# Mid-infrared near-field fingerprint spectroscopy of the 2D electron gas in LaAlO$_3$/SrTiO$_3$ at low temperatures

Julian Barnett[1], Konstantin G. Wirth[1], Richard Hentrich[2], Yasin C. Durmaz[2,3], Marc-André Rose[4], Felix Gunkel[4], and Thomas Taubner[1]*

**Affiliations:**

[1] I. Institute of Physics (IA), RWTH Aachen University, 52074 Aachen, Germany

[2] attocube systems AG, 85540 Haar, Germany,

[3] Department of Physics, Ludwig Maximilians University of Munich, 80799 Munich, Germany

[4] Peter Grünberg Institute (PGI-7) and Jülich-Aachen Research Alliance (JARA-FIT), Forschungszentrum Jülich, 52428 Jülich, Germany

## Abstract

Confined electron systems, such as 2D electron gases (2DEGs), 2D materials, or topological insulators show great technological promise but their susceptibility to defects often results in nanoscale inhomogeneities with unclear origins. Scattering-type scanning near-field optical microscopy (s-SNOM) is useful to investigate buried confined electron systems non-destructively with nanoscale resolution, however, a clear separation of carrier concentration and mobility was often impossible in s-SNOM. Here, we predict a previously inaccessible characteristic "fingerprint" response of the prototypical LaAlO$_3$/SrTiO$_3$ 2DEG, and verify this using a state-of-the-art tunable narrow-band laser in mid-infrared cryo-s-SNOM at 8 K. Our modelling allows us to separate the influence of carrier concentration and mobility on fingerprint spectra and to characterize 2DEG inhomogeneities on the nanoscale. This spatially resolved information about the local electronic properties can be used to identify the origin of inhomogeneities in confined electron systems, making the s-SNOM fingerprint response a valuable tool for nanoelectronics and quantum technology.

## Introduction:

Layered heterostructures of complex oxides have become an integral part of novel electronic, spintronic and magneto-ionic device concepts.[1–4] Such structures often contain high local densities of electronic charge carriers, giving rise to correlation phenomena and emerging properties, such as metallic behavior,[5–7] superconductivity,[8–10] and magnetism,[11–13] not found in the adjacent bulk materials. This poses new challenges to characterize high carrier concentrations in spatially confined and buried electronic systems, and in the typically desired cryogenic temperature range. Scattering-type scanning near-field optical microscopy (s-SNOM)[14,15] is a non-destructive method that uses strong optical near-fields and exhibits high surface sensitivity, sub-surface capabilities, and applicability in a broad spectral range. This makes s-SNOM an ideal candidate to investigate highly-confined electron systems that exist in the transdimensional regime[16] between purely 2D and bulk 3D, such as van-der-Waals materials (e.g. few-layer graphene[17]) or oxide heterostructures (e.g. $LaAlO_3/SrTiO_3$[5]). However, the spectral information that is observed in s-SNOM is convoluted in terms of probing depth and in terms of different physical excitations,[15] such as phonons, plasmons, or interband transitions. This imposes challenges on the qualitative and quantitative understanding of the acquired s-SNOM data, especially in layered structures. For one, the near-field response consists of contributions of multiple layers simultaneously, which makes a detailed understanding and modelling of the near-field response a crucial requirement. Moreover, this calls for the identification of dedicated spectral regions that provide characteristic information on the layer of interest ("fingerprint" regions), adding requirements to the accessible frequency range of light sources at sufficient signal-to-noise ratio.

The 2D electron gas (2DEG) at the interface between the two insulators $LaAlO_3$ (LAO) and $SrTiO_3$ (STO)[5] has been studied extensively as a model system for high-concentration correlated electron systems, however, the local formation process and the influence of defects on the electronic properties are not fully understood.[18–20] S-SNOM was shown to be sensitive to the 2DEG in LAO/STO[21] and even allow for the extraction of local electronic properties with nanoscale lateral resolution.[22–24] Due to the metallic behavior of the system, the 2DEG mobility increases at low temperatures,[5] which should lead to a stronger sensitivity of s-SNOM to the 2DEG properties at cryogenic temperatures (cryo-s-SNOM) compared to room temperature. Previous s-SNOM studies were limited to indirectly probing the 2DEG via secondary effects, i. e. damping of the phonon near-field resonance of the STO substrate[23] or as a constant background at higher frequencies.[22,24] During these studies, direct mapping of local charge carrier density turned out to be difficult, as the influence of different parameters, such as carrier concentration and mobility, on near-field spectra compensates to a certain degree.[24]

In this work, we predict the existence of a spectral fingerprint region, where a characteristic scattering response of the 2DEG can be obtained directly, resulting in a near-field spectroscopic method that allows for the separation of carrier concentration and mobility. This spectral region was previously inaccessible for s-SNOM on LAO/STO due to missing light sources with sufficient signal-to-noise ratio, as the overall scattering signal of the fingerprint region is very low. Here, we use a newly-developed tunable narrow-band mid-infrared laser to investigate LAO/STO in the 2DEG fingerprint region, using cryo-s-SNOM at 8 K. This allows us to use a normalization procedure that is highly sensitive to the influence of the 2DEG and can be used to characterize the local electronic properties in detail.

Figure 1a presents a simplified view of the experimental setup (cf. Methods for details), with the metal-coated AFM tip scanning across the LAO/STO sample while oscillating with tapping frequency $\Omega_{tip}$ and being illuminated with a focused laser beam. The measurements presented in this work were recorded with a "cryo-neaSCOPE" by attocube systems AG, with the samples being at a constant temperature of 8 K. For illumination, the narrow-band (linewidth < 5 cm$^{-1}$) laser "PT277-XIR" by EKSPLA was used, which is based on a picosecond optical parametric oscillator with difference frequency generation and exhibits a high signal-to-noise ratio in the fingerprint region of the LAO/STO 2DEG. While the spot size of the focused laser on the sample is diffraction limited to $\lambda/2$ (several micrometers in mid-IR), the near-fields at the tip apex depend on the tip radius and enable a lateral resolution down to the ten-nanometer range. The back-scattered light is demodulated at higher harmonics $n\Omega_{tip}$ of the tapping frequency and pseudo-heterodyne interferometric detection allows for separation into near-field scattering amplitude $s_n$ and phase $\varphi_n$. In s-SNOM, the absolute scattering signal strongly depends on the experimental setup, e. g. the tip shape, oscillation amplitude, detector sensitivity, or laser power. To acquire setup-independent results, it is an established technique to use normalization to a reference sample, resulting in the relative scattering amplitude $s_n/s_n^{ref}$ and phase $\phi_n-\phi_n^{ref}$.

This near-field scattering amplitude and phase depends on the frequency-dependent optical properties of the sample, expressed as the dielectric function or permittivity $\varepsilon(\omega)$. Figure 1b shows the real (solid lines) and imaginary (dashed lines) part of the permittivity of bulk STO (blue) and bulk LAO (red), both of which exhibit phonon resonances in the mid-IR range. As the near-fields of s-SNOM decay exponentially with distance to the tip, even layers of few nanometers thickness can significantly influence the scattering signal,[25–27] especially in the vicinity of zero-crossings of the materials' $Re[\varepsilon]$.[23,28,29] At the same time, near-fields typically still have a penetration depth of several ten nanometers,[30–32] leading to a combined scattering response dependent on the permittivities of the STO substrate, the LAO top layer, and the 2DEG in between. To isolate the influence of the 2DEG layer from the contributions of the LAO covering layer and STO substrate, a direct comparison of samples with and without 2DEG, respectively, is helpful, ideally showing identical lattice phonon properties. This can be achieved by controlling the termination of the STO substrate (Figure 1c): SrO-terminated STO surfaces lead to an insulating interface (without 2DEG) after LAO deposition, while $TiO_2$-terminated surfaces lead to a conductive interface (with 2DEG), if a critical thickness of at least 4 unit cells (uc) of LAO is deposited epitaxially.[5,24] The samples used in this work were fabricated by pulsed laser deposition (PLD) of 8 uc (3 nm) LAO on single crystal STO substrates. All substrates were wet-etched to get pure $TiO_2$ termination and then for one of them a single SrO layer was applied by PLD, to change the termination (cf. Methods for details).[33] Comparing these two types of LAO/STO samples allows for a direct investigation of the 2DEG response at low temperatures by using the non-conducting interface as a reference in the s-SNOM measurement.

To showcase the influence of the different layers, Figure 1d presents simulations of the gold-normalized near-field scattering amplitude $s_2/s_2^{Au}$ for bulk STO (blue), LAO/STO without 2DEG (violet) and LAO/STO with 2DEG (dashed orange), at room temperature (cf. reference [23]). For all simulations and measurements, the demodulation order $n=2$ is presented here, to get a high signal-to-noise ratio with sufficient background suppression in the observed spectral range. Simulations were done using the finite dipole model[34] for the near-field calculations, combined with the transfer matrix method to get the near-field response of an arbitrary layer stack (cf. Methods for details).[35] The 2DEG was modelled as several layers of exponentially decaying carrier density,[22] in accordance with depth profile assumptions from literature.[36,37] The STO substrate alone (blue) shows a strongly enhanced scattering amplitude ("near-field resonance") slightly below its LO-frequency $\omega_{LO}=788$ cm$^{-1}$, where $Re[\varepsilon_{STO}]$ is slightly negative and $Im[\varepsilon_{STO}]$ is low. When adding the ultra-thin LAO on top of STO (violet), the overall shape of the STO near-field resonance persists but is modified by the zero-crossings of $Re[\varepsilon_{LAO}]$ around 600, 650, and 750 cm$^{-1}$, resulting in amplitude features with derivative line shape. Adding 2DEG charge carriers at the interface (dashed orange), the scattering amplitude of the near-field resonance around 700 cm$^{-1}$ is reduced by additional damping and the high-frequency scattering above 900 cm$^{-1}$ is slightly increased (barely visible at this scale). Grey-shaded areas indicate spectral regions where the 2DEG was investigated in previous studies. These studies were limited to the influence of the 2DEG on i) the phonon near-field resonance of STO (< 750 cm$^{-1}$), where the s-SNOM signal is generally high,[23] or II) the off-resonance scattering response in the spectral window of light sources with a high signal-to-noise ratio, such as $CO_2$ lasers (> 920 cm$^{-1}$).[22,24] In these two regions, the general influence of free charge carriers on near-field spectra is only indirect, either as an additional source of damping (increased Im[$\varepsilon$]) or as a constant background (increased high-frequency limit $\varepsilon_\infty$, cf. Supplementary Information).

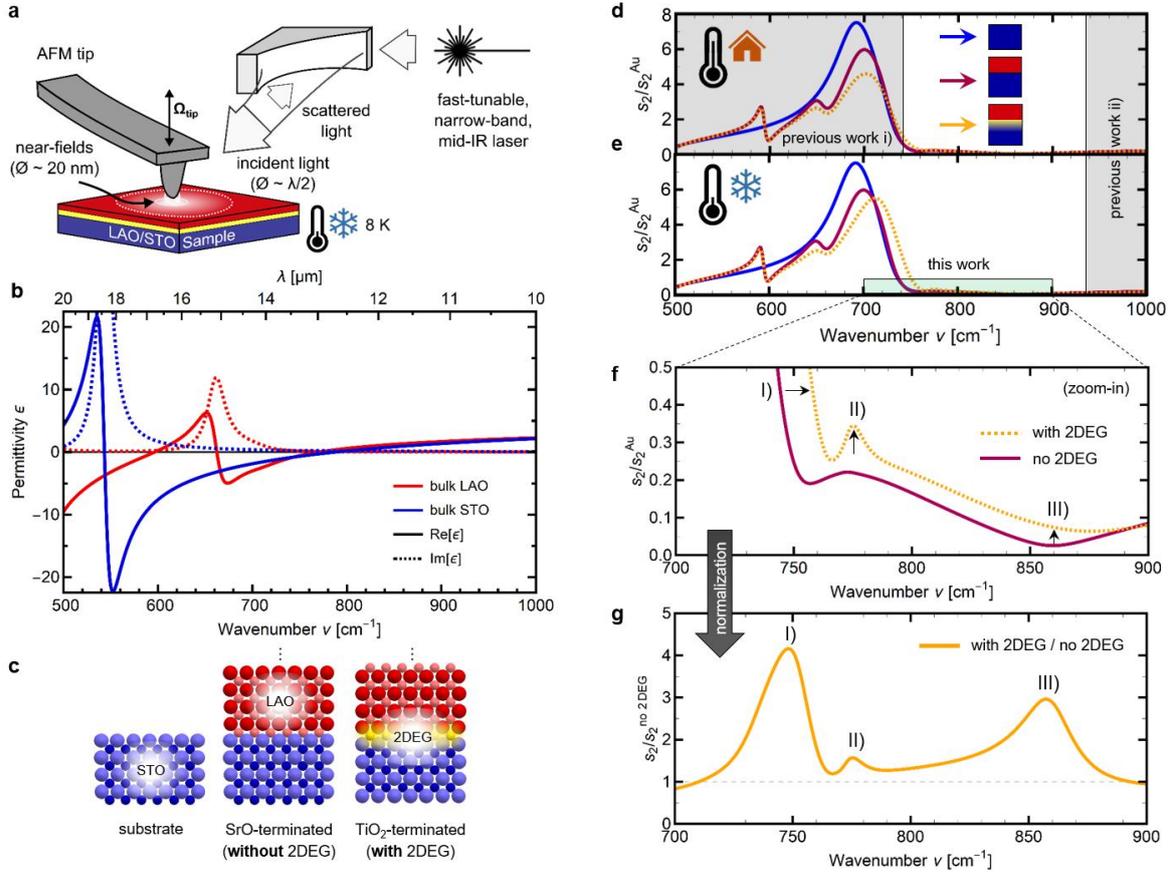

Figure 1: Investigation of LaAlO$_3$/SrTiO$_3$ (LAO/STO) interfaces using scanning near-field optical microscopy (s-SNOM) and spectroscopy in the 2DEG fingerprint region. **a** Experimental setup showing the LAO/STO sample (cooled to 8 K) and the metal-coated AFM tip oscillating at tapping frequency $\Omega_{tip}$, illuminated by a new fast-tunable narrow-band mid-infrared laser. **b** Real (solid lines) and imaginary part (dashed lines) of the dielectric functions of STO (blue) and LAO (red). Both oxides show phonon resonances in the mid-infrared range. **c** STO single crystals (left) were epitaxially covered with 8 uc of LAO. Depending on the termination of the STO substrate, the interface is either insulating (SrO-terminated) or conducting (TiO$_2$-terminated), the latter hosting a 2D electron gas (2DEG). **d** Simulation of the Au-normalized near-field amplitude spectra $s_2/s_2^{Au}$ for STO substrate (blue), insulating LAO/STO (violet) and LAO/STO with 2DEG (dashed orange) at room temperature. Grey-shaded areas indicate spectral regions that have been investigated at room temperature before.[22–24] **e** Simulation of $s_2/s_2^{Au}$ at cryogenic temperatures (with increased 2DEG mobility). The shaded areas indicate the spectral range investigated at low temperatures previously[22] (grey) and in this work (green). **f** Zoom-in of Figure 1d to low scattering amplitudes in the fingerprint region, showing the insulating (violet) and conducting (dashed orange) LAO/STO samples. **g** By referencing the conducting to the insulating case (dividing of scattering amplitude), changes introduced by the 2DEG are highlighted, resulting in a characteristic response with three separate maxima (fingerprint spectrum).

## Results

Figure 1e shows how the near-field spectrum of the conducting LAO/STO interface (dashed orange) is expected to change at low temperatures, where the higher 2DEG mobility leads to a slight shift of the peak to higher frequencies and an increased scattering amplitude. Zooming-in to frequencies above the phonon near-field resonance, Figure 1f shows significant differences between the scattering amplitude of the SrO-terminated (without 2DEG, violet) and the TiO$_2$-terminated LAO/STO interface (with 2DEG, dashed orange). Adding the 2DEG leads to three changes in the spectrum, as indicated by black arrows: I) a shift of the high-frequency flank of the STO phonon near-field resonance to higher frequencies, II) an additional spectral feature around 775 cm$^{-1}$, and III) a general increase in scattering amplitude. However, the overall s-SNOM signal in the spectral range above 750 cm$^{-1}$ is quite low, with relative scattering amplitudes between 2% and 30% of that of gold (cf. Figure 1d/e), and thus a high signal-to-noise ratio of the illumination source is necessary for near-field spectroscopy. If the signal-to-noise ratio in this highly characteristic frequency range is sufficiently high, a normalization procedure between samples with and without 2DEG, respectively, can be employed, which is very sensitive to the 2DEG properties.

Figure 1g showcases the theoretical prediction of this characteristic 2DEG near-field response, that can be obtained by referencing the sample with 2DEG to the sample without 2DEG, i. e. by using Figure 1e and dividing the data plotted in dashed orange by that in solid violet. The resulting normalized amplitude spectrum shows only changes to the spectrum induced by adding the 2DEG charge carriers to the non-conducting interface, with three distinct peaks that relate to the aforementioned changes (I, II, III), introduced in Figure 1e. As the height, position and shape of these features are very sensitive to changes of the electronic properties (cf. Figure 2), they can be seen as the characteristic "fingerprint" near-field spectrum of this 2D electron system, similar to characteristic vibrational bands of polymers in near-field spectroscopy.[38] We have thus identified a unique fingerprint signature of the LAO/STO 2DEG, with three characteristic peaks in the spectral range of 700 to 900 cm$^{-1}$, constrained to this frequency range by the phonon properties of the two materials.

To illustrate how fingerprint spectra of the 2DEG (cf. Figure 1g) are influenced by its electronic properties, Figure 2 shows theoretical predictions of the changes introduced by a varying 2DEG mobility $\mu$ (Figure 2a) and 2DEG sheet carrier concentration $n_{2D}$ (Figure 2b). Values for carrier concentration and mobility were picked in accordance with Hall measurements of the sample, with experimental values of $n_{2D} = 4.3 \times 10^{13}$ cm$^{-2}$ and $\mu = 5.1$ cm²/Vs at room temperature. While the 2DEG mobility can increase to above $10^4$ cm²/Vs upon cooling,[5] it should be noted that the effective mobility in the mid-infrared spectral range can be significantly lower than those measured in the far-infrared spectral range or in transport measurements, due to spectral weight redistribution from polaronic contributions.[37,39] In Figure 2a, a rising mobility (transition from dark red to yellow) from 5 to 80 cm²/Vs leads to an increase of the relative scattering amplitude for the peak around 750 cm$^{-1}$, while the other two peaks remain mostly unchanged. Contrary to that, a rise in carrier concentration (transition from dark red to yellow in Figure 2b) leads to an increase of the relative scattering amplitude for all three peaks. Additionally, the scaling behaviour is very different (cf. Figure S1 in the Supplementary Information). Even though μ is doubled between each curve in Figure 2a, the increase in peak height diminishes with each step and saturation is reached around 80 cm²/Vs (yellow curve). In comparison, $n_{2D}$ is only increased linearly and the increase in peak height of the relative scattering signal persists across the whole value range shown here.

Figure 2c presents the first ever s-SNOM measurements of the 2DEG fingerprint region, for two different positions (black, grey) on LAO/STO with 2DEG (cf. Methods and Supplementary information for details). For the first position (black curve), a high peak (I) around 750 cm$^{-1}$, a small peak (II) at 790 cm$^{-1}$ and a broad peak (III) around 850 cm$^{-1}$ can be observed, which fits well to the theoretical predictions presented in Figures 2a/b. The fingerprint spectrum recorded at the second position (grey) deviates significantly from the first, with a lower intensity in the first (I), a higher intensity in the second (II), and a less defined shape in the third peak (III). Additionally, the minimum around 775 cm$^{-1}$ and the maximum around 850 cm$^{-1}$ are both shifted to lower frequencies by 5-10 cm$^{-1}$. Figure 2d presents simulations with different parameter sets, to reproduce the measurements from Figure 2c: set A (black) was achieved with $n_{2D} = 6 \times 10^{13}$ cm$^{-2}$ and $\mu = 20$ cm²/Vs, while set B (grey) relates to $n_{2D} = 8 \times 10^{13}$ cm$^{-2}$ and $\mu = 10$ cm²/Vs. Generally, good agreement between 2DEG fingerprint spectra and simulations can be achieved, showing that the theoretical description of the 2DEG with finite dipole model, transfer matrix method, exponentially decaying depth profile, and phonon background represents the physical behaviour of the system well. The first parameter set (black) reproduces the first measurement position (black in Figure 2c) well for the first (I) and second (II) peak, while the third peak is less broad and lower in intensity. The second parameter set (grey) reproduces key features of the second measurement position (grey in Figure 2c): the different intensity distribution between the first two peaks (I) and (II) and the shift of the minimum around 775 cm$^{-1}$ and the third maximum (III) to lower frequencies. However, this matching was not possible by changing the electronic properties alone. Instead, small changes to the LAO/STO phonon properties were necessary to achieve the frequency shift of the minimum (cf. Supplementary Information) and the third maximum (III). This points towards inhomogeneities of the LAO layer, possibly due to differently strained areas after cooling, as strain can lead to significant shift of the phonon frequencies in oxides,[40] while inhomogeneous 2DEG properties could result from magnetic or structural domains observed for LAO/STO 2DEGs at low temperature.[41–44] The higher background at 900 cm$^{-1}$ in both measurements compared to the respective simulations could be explained by a change in the high-frequency limit $\varepsilon_\infty$ of the dielectric function, that was shown to contribute to STO near-field signals upon doping.[45,46]

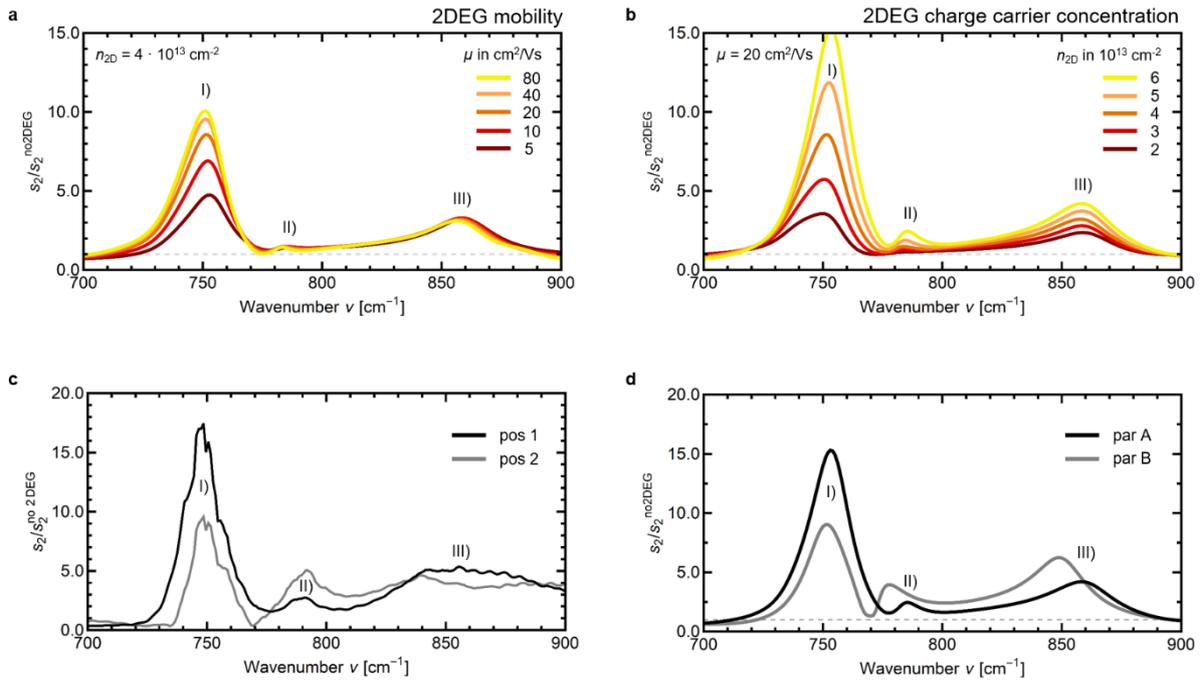

Figure 2: Fingerprint spectra of the 2DEG at conducting LAO/STO interfaces, normalized to the non-conducting interface (cf. Figure 1f, g). **a** Simulation for 2DEG mobilities $\mu$ between 5 cm²/Vs (dark red) and 80 cm²/Vs (yellow), with a constant 2DEG carrier concentration $n_{2D} = 4\cdot 10^{13}$ cm$^{-2}$. **b** Simulation for different $n_{2D}$ between $2\cdot 10^{13}$ cm$^{-2}$ (dark red) and $6\cdot 10^{13}$ cm$^{-2}$ (yellow), with a constant $\mu = 20$ cm²/Vs. **c** Measurement of fingerprint spectra at two different positions of the conducting sample. **d** Simulation of fingerprint spectra with different simulation parameter sets (cf. Supplementary Information), reproducing the experimental spectra shown in panel c.

To further investigate these differences between different positions on the sample, s-SNOM can be used to record lateral images of conducting LAO/STO surfaces on the micro-/nanoscale, which is presented in Figure 3a for the scattering amplitude $s_2$ at a frequency of 700 cm$^{-1}$, where the phonon near-field response is strongest (cf. Figure 1d/e). In particular, single frequency s-SNOM images were recorded, that show significant variations in amplitude $s_2$ (Figure 3a) and phase $\phi_2$ (Supplementary Information,

Figure S2). This local inhomogeneity was further investigated with near-field spectra at measurement positions 1-5 indicated by crosses in Figure 3a. The recorded spectra were again normalized to the insulating interface and the results are presented in Figure 3b, showing the previously discussed fingerprint spectrum of the 2DEG, with a high (I), a small (II), and a broad peak (III). Interestingly, moving from measurement position 1 (dark blue) to 5 (yellow), the intensity of the first peak (I) increases by 50%, while the other two peaks remain mostly unchanged. This behaviour is very similar to Figure 2a and can be well reproduced by a variation of the 2DEG mobility $\mu$ between 8 cm²/Vs and 40 cm²/Vs (Figure 3c). This indicates that the local inhomogeneities observed in the s-SNOM image (Figure 3a) could be explained by a laterally varying 2DEG mobility and showcases the sensitivity of 2DEG fingerprint spectra. Additional s-SNOM images recorded after the line scan (cf. Supplementary Information, Figure S2) show that the measurement itself could influence the local electronic properties, possibly due to electrostatic gating from the irradiated AFM tip,[22] persistent photoconductivity,[47,48] or frozen condensates.[49] This effect becomes increasingly relevant as the sensitivity of the method to the local properties increases and presents an important avenue for future research.

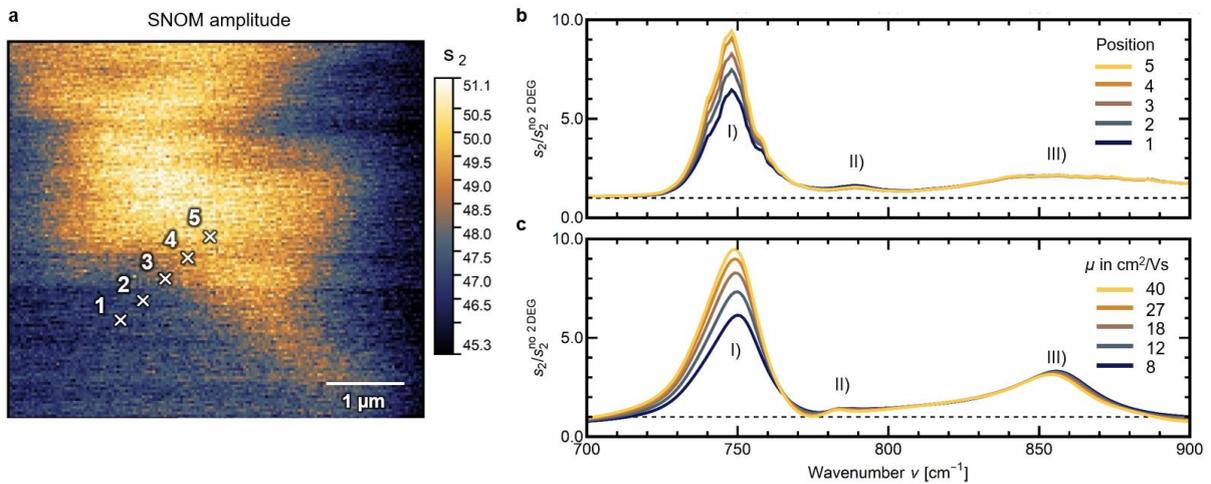

Figure 3: Fingerprint investigation of lateral inhomogeneities on the nanoscale. **a** S-SNOM image of near-field amplitude $s_2$ at $\omega = 700$ cm⁻¹, close to the maximum of the scattering response from the STO phonon near-field resonance (cf. Figure 1b-d). Markers 1-5 indicate measurement positions of a line scan with the resulting spectra shown in Figure 3b. **b** Fingerprint spectra measured at positions 1 (blue) to 5 (yellow) of Figure 3a. **c** Simulation of fingerprint spectra with 2DEG mobility $\mu$ rising from 8 cm²/Vs (blue) to 40 cm²/Vs (yellow).

## Conclusion & Outlook

To conclude, we presented theoretical predictions of a characteristic spectral region for the s-SNOM investigation of the LAO/STO 2DEG at low temperatures. In this region, s-SNOM fingerprint spectra of the 2DEG can be recorded by normalizing samples of different substrate termination. These fingerprint spectra are highly sensitive to changes of the electronic properties, even though the overall s-SNOM signal in this spectral region is very low. Three characteristic peaks can be found, which are differently influenced by 2DEG carrier concentration and mobility, allowing for their disentanglement from spectroscopic measurements. The changes introduced by rising 2DEG mobility mostly relate to the main fingerprint peak around 750 cm⁻¹, with a saturation of the influence around $\mu = 80$ cm²/Vs. The influence of the charge carrier density is stronger by comparison and influences all three peaks similarly. Using a newly-developed light source in combination with cryo-s-SNOM at 8 K, we were able to record s-SNOM fingerprint spectra of the 2DEG in this previously inaccessible spectral window with pseudo-heterodyne point spectroscopy, due to the high signal-to-noise ratio. We found local variations in measured fingerprint spectra, which could be explained with changes of the 2DEG and phonon properties.

In the future, this can be used to disentangle the local carrier concentration and mobility in investigations of the 2DEG in LAO/STO, for example for the characterization of different types of defects,[50] such as varying LAO thickness, atomic steps, or vacancies of different ions. Additionally, the higher 2DEG mobility at low temperature increases the surface sensitivity of s-SNOM compared to room temperature and should allow for conclusions about the vertical distribution[37,51] of the charge carriers. More generally, novel light sources allow fast frequency sweeping across a broad spectral range in the infrared region with a high signal-to-noise ratio, which enables normalization between samples with similar properties in near-field spectroscopy. This normalization makes s-SNOM very sensitive to small changes in the optical properties, as was successfully shown for interband transitions in tetralayer graphene recently,[52] even when overall s-SNOM signals are very low. Upon identification of promising characteristic frequencies, s-SNOM images can be recorded using the narrow linewidth,[45] to investigate local defects with nanoscale resolution at the frequency where the contrast is strongest. Here, we have demonstrated a link between quantitative analysis and spatially resolving carrier density and mobility in the 2DEG of LAO/STO via fingerprint spectra. This method can be easily transferred to other confined correlated electron systems such as 2DEGs at different (oxide) heterointerfaces,[53,54] van-der-Waals heterostructures or multilayers of 2D materials,[52,55] or topological insulators,[56,57] providing spatially resolved information for designing nanoelectronic devices.

# Methods

1. Sample preparation

Eight unit cells of $LaAlO_3$ were deposited on wet-etched, $TiO_2$-terminated (100)-$SrTiO_3$ substrates using pulsed laser deposition (PLD). The films were deposited at a laser frequency of 1 Hz at a fluence of 1.0 J/cm$^2$ and an oxygen pressure of $1\times10^{-4}$ mbar. The growth temperature was 800°C and the samples were quenched down to room temperature after a relaxation at growth temperature.[58] During growth, clear RHEED-intensity oscillations were observed indicating a layer-by-layer growth mode and yielding single unit cell thickness control. The samples were characterized by AC-Hall effect measurements in van-der-Pauw geometry at room temperature indicating a sheet carrier density of $4.3\times10^{13}$ cm$^{-2}$ and an electron mobility of 5.1 cm$^2$/Vs. SrO termination was achieved by PLD deposition in the same system as for the LAO thin films from a ceramic $SrO_2$ target using a laser fluence of 0.9 J/cm$^2$. The chamber was held at an oxygen partial pressure $2\times10^{-7}$ mbar and the substrate was heated to 800°C. To achieve the change of termination, the deposition was stopped as soon as the first order diffracted spot fulfilled one oscillation, crossing the intensity of the specular spot.[59]

2. S-SNOM measurements

S-SNOM measurements were done using a commercial low-temperature scattering-type scanning near-field optical microscope (cryo-neaSCOPE by attocube systems AG), with a pseudo-heterodyne detection configuration[60] for single-frequency 2D imaging and sequential point spectroscopy. As light source, a commercial narrow-band tunable laser (ekspla PT277-XIR) was used, consisting of a tunable optical parametric oscillator (OPO) combined with difference frequency generation (DFG), that delivers beams of approx. 10 mW power in the DFG region (625-2000 cm$^{-1}$), in 8 ps pulses with 87 MHz repetition rate and a typical line width of $< 3$ cm$^{-1}$ in this frequency range. As a detector, a photoconductive Mercury Cadmium Telluride (HgCdTe) detector element of 50 µm diameter with a ZnSeW window was used. As scattering tips, commercial nano-FTIR probes by attocube systems AG were used, with a tapping amplitude of approximately 80 nm. For in-situ referencing, both samples (with and without 2DEG, respectively) were glued side by side on a copper plate mounted on the sample stage. The system was cooled down to its base temperature with the samples being stabilized at $T = 8$ K, as monitored by a calibrated Cernox® sensor which was thermally coupled to the sample plate. Cooling was provided by an integrated closed cycle cryostat (attoDry800 from attocube), keeping the sample space free of cryogenic media.

3. Theoretical Modelling

The Finite Dipole Model (FDM)[34] and Transfer Matrix Method (TMM) were used in combination[35] to determine the near-field scattering response of arbitrarily layered systems with known dielectric functions. FDM parameters were 400 nm ellipsoid length, 90 nm tip radius, 80 nm tapping amplitude, demodulation order $n = 2$, and the geometric factor $g = 0.7\times\exp(0.1i)$. The $p$-polarized TMM reflection coefficient was used as the FDM sample reflection factor,[35] at a dominant in-plane wavevector[26] of $k_x = 250\,000$ cm$^{-1}$. To model the optical properties of the layer stack, the dielectric functions of STO and LAO were calculated from literature data[61,62] using the Berreman-Unterwald-Lowndes factorized form[63,64] of multiple Lorentz oscillators (cf. Supplementary Information). To describe the dielectric function of the 2DEG, a Drude term was added to the STO phonon background,[64] which was then used in a multilayer approach of 10 slices with thickness $d = 1$ nm and exponentially decaying charge carrier concentration away from the interface, distributing the 2D sheet carrier density with a decay constant $z_0 = 2$ nm. The effective mass was approximated to an averaged effective mass of $m^* = 3.2$ $m_0$ (with the free electron mass $m_0$) as predicted by First Principles calculations[51] and measured experimentally.[37]

## Acknowledgements

J.B., K.W., and T.T. acknowledge funding from the Deutsche Forschungsgesellschaft (DFG) within the collaborative research center SFB 917 and under grant agreement no. TA 848/7-1. M.A.R. and F.G. acknowledge funding financial support from the Deutsche Forschungsgesellschaft (DFG) FG 1604 (No. 315025796).

## Author Contributions

J.B., T.T., and F.G. conceived and planned the study. J.B., Y.C.D. and R.H. planned and prepared the measurements, which J.B., T.T., Y.C.D, and R.H. carried out. M.A.R. and F.G. provided the samples; J.B., M.A.R. and F.G. did pre- and post-characterization. J.B. and K.W. performed data evaluation and simulations. All authors discussed the data and contributed to writing the paper. All authors approved the final version of the manuscript.

## Competing Interests

Y.C.D. and R.H. are employees of attocube systems AG, producer of the cryo-s-SNOM microscope used in this study.

# Mid-infrared near-field fingerprint spectroscopy of the 2D electron gas in LaAlO$_3$/SrTiO$_3$ at low temperatures

Julian Barnett[1], Konstantin G. Wirth[1], Richard Hentrich[2], Yasin C. Durmaz[2,3], Marc-André Rose[4], Felix Gunkel[4], and Thomas Taubner[1]*

**Affiliations:**

[1] I. Institute of Physics (IA), RWTH Aachen University, 52074 Aachen, Germany
[2] attocube systems AG, 85540 Haar, Germany,
[3] Department of Physics, Ludwig Maximilians University of Munich, 80799 Munich, Germany
[4] Peter Grünberg Institute (PGI-7) and Jülich-Aachen Research Alliance (JARA-FIT), Forschungszentrum Jülich, 52428 Jülich, Germany

## Supplementary Information

1. <u>Simulation parameters</u>

**Figure 1b and following:** The bulk dielectric functions of LaAlO$_3$ (LAO) and SrTiO$_3$ (STO) were calculated using the Berreman-Unterwald-Lowndes factorized form:[1,2]

$$\varepsilon_j^L(\omega) = \varepsilon_{\infty,j} \prod_l^k \frac{\omega^2 + i\gamma_{\text{LO},lj}\omega - \omega_{\text{LO},lj}^2}{\omega^2 + i\gamma_{\text{TO},lj}\omega - \omega_{\text{TO},lj}^2}$$

with high-frequency limit $\varepsilon_\infty$, transverse optical (TO-) and longitudinal optical (LO-) frequencies $\omega_{\text{TO/LO}}$, and respective damping parameters $\gamma_{\text{TO/LO}}$ taken from literature:[3,4]

*Table 1: literature data for the dielectric function of LAO (all values in cm$^{-1}$):*

| $\omega_{\text{TO},1}$ | $\gamma_{\text{TO},1}$ | $\omega_{\text{LO},1}$ | $\gamma_{\text{LO},1}$ | $\omega_{\text{TO},2}$ | $\gamma_{\text{TO},2}$ | $\omega_{\text{LO},2}$ | $\gamma_{\text{LO},2}$ | $\omega_{\text{TO},3}$ | $\gamma_{\text{TO},3}$ | $\omega_{\text{LO},3}$ | $\gamma_{\text{LO},3}$ |
|---|---|---|---|---|---|---|---|---|---|---|---|
| 188.0 | 0.4 | 276.4 | 3.7 | 427.0 | 5.0 | 596.1 | 7.2 | 495.7 | 3.8 | 495.5 | 3.8 |
| $\omega_{\text{TO},4}$ | $\gamma_{\text{TO},4}$ | $\omega_{\text{LO},4}$ | $\gamma_{\text{LO},4}$ | $\omega_{\text{TO},5}$ | $\gamma_{\text{TO},5}$ | $\omega_{\text{LO},5}$ | $\gamma_{\text{LO},5}$ | $\varepsilon_{\text{st}}$ | $\varepsilon_\infty$ | | |
| 650.8 | 22.5 | 744.1 | 12.1 | 708.2 | 55.3 | 702.2 | 66.0 | 22.3 | 4.12 | | |

*Table 2: literature data for the dielectric function of STO (all values in cm$^{-1}$):*

| $\omega_{\text{TO},1}$ | $\gamma_{\text{TO},1}$ | $\omega_{\text{LO},1}$ | $\gamma_{\text{LO},1}$ | $\omega_{\text{TO},2}$ | $\gamma_{\text{TO},2}$ | $\omega_{\text{LO},2}$ | $\gamma_{\text{LO},2}$ | $\omega_{\text{TO},3}$ | $\gamma_{\text{TO},3}$ | $\omega_{\text{LO},3}$ | $\gamma_{\text{LO},3}$ | $\varepsilon_{\text{st}}$ | $\varepsilon_\infty$ |
|---|---|---|---|---|---|---|---|---|---|---|---|---|---|
| 91 | 15.0 | 172 | 3.8 | 175 | 5.4 | 474 | 4.5 | 543 | 17.0 | 788 | 25 | 310 | 5.1 |

**Figure 1c, 1d, 1f, and following:**

Finite dipole model (FDM) parameters were 400 nm ellipsoid length, 90 nm tip radius, 80 nm tapping amplitude, demodulation order $n = 2$, and the geometric factor $g = 0.7 \times \exp(0.1i)$. The *p*-polarized transfer matrix method (TMM) reflection coefficient was used as the FDM sample reflection factor,[5] at a dominant in-plane wavevector[6] of $k_x = 250\,000$ cm$^{-1}$. The LAO thickness was 8 unit cells (3 nm) and the dielectric function of the 2D electron gas (2DEG) was described by adding a Drude term to the phonon background of STO:[2]

$$\varepsilon_j^{DL}(\omega) = \varepsilon_j^L(\omega) + \varepsilon_j^D(\omega) = \varepsilon_{\infty,j} \prod_l^k \frac{\omega^2 + i\gamma_{LO,lj}\omega - \omega_{LO,lj}^2}{\omega^2 + i\gamma_{TO,lj}\omega - \omega_{TO,lj}^2} - \frac{\omega_p^2}{\omega^2 + \gamma^2} + i\frac{\gamma\omega_p^2}{\omega(\omega^2 + \gamma^2)}$$

Here, $\omega_p$ is the plasma frequency and $\gamma$ is the electron damping parameter, respectively described as:

$$\omega_p^2 = \frac{n_{2D}}{z_0} \cdot \exp\left(-\frac{z}{z_0}\right) \cdot \frac{e^2}{\varepsilon_0 m^*} \quad \text{and} \quad \gamma = \frac{e}{m^* \mu}$$

for an exponentially decaying carrier concentration.[7] To model the depth distribution of the 2DEG, $z$ is the distance from the LAO/STO interface, $\mu$ is the carrier mobility and $n_{2D}$ the sheet carrier density. The latter is transformed into a volume carrier density which then decays exponentially with decay constant $z_0$ away from the interface. This distance-dependent dielectric function was then used in a multilayer approach of 10 slices with thickness $d = 1$ nm, with an averaged effective mass of $m^* = 3.2$ $m_0$. For Figure 1d, the values are $\mu = 2$ cm²/Vs and $n_{2D} = 3\times10^{13}$ cm⁻², for Figure 1e and 1f the mobility is enhanced to $\mu = 10$ cm²/Vs.

**Figure 2:**

*Table 3: modelling parameters used in Figure 2, compared to Hall measurement data of the sample:*

|  | $n_{2D}$ (cm⁻²) | $\mu$ (cm²/Vs) |
|---|---|---|
| Hall meas. (T = 300 K) | $4.3 \times 10^{13}$ | 5.1 |
| Figure 2a | $4 \times 10^{13}$ | 5, 10, 20, 40, 80 |
| Figure 2b | $(2, 3, 4, 5, 6) \times 10^{13}$ | 20 |
| Figure 2c, *parameter set A* | $6 \times 10^{13}$ | 20 |

*Table 4: modelling parameters used in Figure 2c, parameter set B (variation of phonon parameters):*

|  | $n_{2D}$ (cm⁻²) | $\mu$ (cm²/Vs) | $\omega_{TO}$ (cm⁻¹) | $\omega_{LO}$ (cm⁻¹) | $\gamma_{TO}$ (cm⁻¹) | $\gamma_{LO}$ (cm⁻¹) |
|---|---|---|---|---|---|---|
| **2DEG** | $8 \times 10^{13}$ | 10 |  |  |  |  |
| **LAO** |  |  | 660 | 780 | 22.5 | 8 |
| **STO** |  |  | 570 | 785 | 17 | 25 |

**Figure 3c:**

*Table 5: modelling parameters used in Figure 3c:*

| $n_{2D}$ (cm⁻²) | $\mu$ (cm²/Vs) |
|---|---|
| $4 \times 10^{13}$ | 8, 12, 18, 27, 40 |

2. Measurement parameters

**For all measurements:**
  frequency resolution: 1 cm$^{-1}$
  PsHet modulation frequency: 300 Hz
  sample temperature: 8 K

**Figure 2c:**

*Table 6: measurement parameters used in Figure 2c:*

| parameter | pos 1 | pos 2 |
|---|---|---|
| frequency (cm$^{-1}$) | 600 – 950 | 600 – 950 |
| averaging | 2 | 1 |
| integration time (ms) | 98 | 98 |
| tip frequency (kHz) | 237 | 237 |
| tapping amplitude (nm) | 77 | 80 |

**Figure 3a / S2:**

*Table 7: measurement parameters used in Figures 3a and S2:*

| parameter | before linescan | linescan | after linescan |
|---|---|---|---|
| scan size (μm) | 5.1 × 5.1 | 5.6 | 5.1 × 5.1 |
| scan resolution (px) | 150 × 150 | 20 | 150 × 150 |
| frequency (cm$^{-1}$) | 700 | 600 – 950 | 700 |
| averaging | 1 | 2 | 1 |
| integration time (ms) | 20 | 98 | 20 |
| tip frequency (kHz) | 237 | 237 | 237 |
| tapping amplitude (nm) | 77 | 85 | 85 |

3. Behavior of peak height with mobility and carrier concentration

The discussion of Figure 2 of the main text mentioned the different scaling behavior of fingerprint spectra with 2DEG mobility and 2DEG carrier concentration, respectively (Figure S1 reproduces the fingerprint spectra from Figure 2a and 2b). To quantify this scaling behavior, Figures S1c and S1d present the maximum height of the fingerprint peak (I) at 750 cm$^{-1}$, in dependence of 2DEG mobility and 2DEG carrier concentration, respectively. It can be seen that the peak height saturates around $s_2/s_2^{\text{no2DEG}} = 10$ with rising mobility, while a slightly superlinear behavior can be observed with rising carrier concentration.

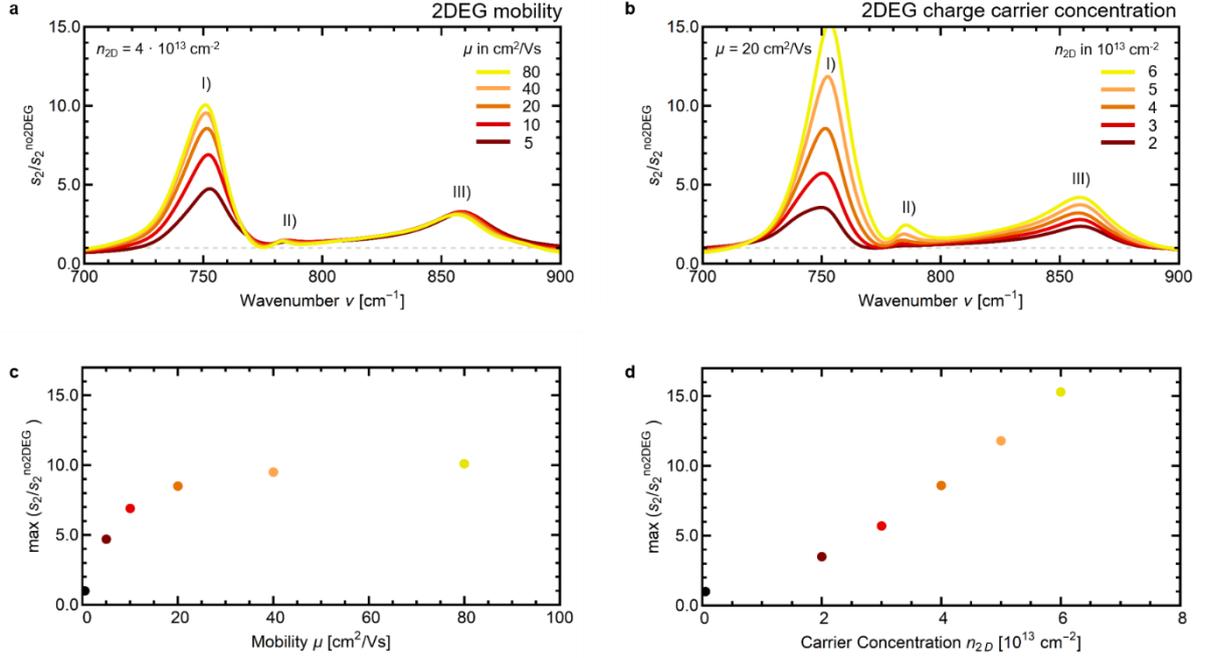

Figure S1: Scaling behavior of fingerprint spectra with electronic properties. Dependence of fingerprint spectra on **a** 2DEG mobility and **b** carrier concentration, reproduced from Figure 2 of the main text. Influence of **c** 2DEG mobility and **d** carrier concentration on maximum height of fingerprint peak (I) at 750 cm$^{-1}$. The black data points in **c** and **d** demark the contrast of $s_2/s_2 = 1$ for the insulating LAO/STO interface.

4. Inhomogeneity and influence of line scan

Figure S2 presents scattering-type scanning near-field optical microscopy (s-SNOM) images of the conducting LAO/STO sample recorded at $\nu = 700$ cm$^{-1}$, where the overall near-field signal is high due to the phonon near-field resonance of STO. As explained in the discussion of Figure 3 of the main text, an inhomogeneity observed in the near-field amplitude (Figure S1a) and the corresponding phase (Figure S1b) were further investigated with a hyperspectral line scan along the dashed line, i. e. sequential spectroscopy by varying the laser frequency between 600 and 950 cm$^{-1}$ at each point along the line. After recording this line scan, another s-SNOM image was recorded in amplitude (Figure S2d) and phase (Figure S2e). While the topography (Figure S2c and S2f) shows that the measurement position was unchanged (indicated by red ovals), the optical image changed drastically in both amplitude and phase. In Figure S2e, a lower phase value (dark blue) is visible along the diagonal from bottom left to top right, following the position of the line scan. This could indicate that the s-SNOM measurement itself is responsible for change in near-field response, possibly due to changes in the local potential (cf. Main Text for discussion). This is further supported by the fact that the first image (top row) was aborted after ~90% completion (black region at the lower end), which can be seen in the second image (bottom row) as a bright yellow square in the amplitude and a blue square in the phase. Therefore, it can be assumed that the inhomogeneity observed in the first image is itself a result of the preprogrammed movement algorithm of the scanning tip, that is executed when starting and aborting measurements from different positions.

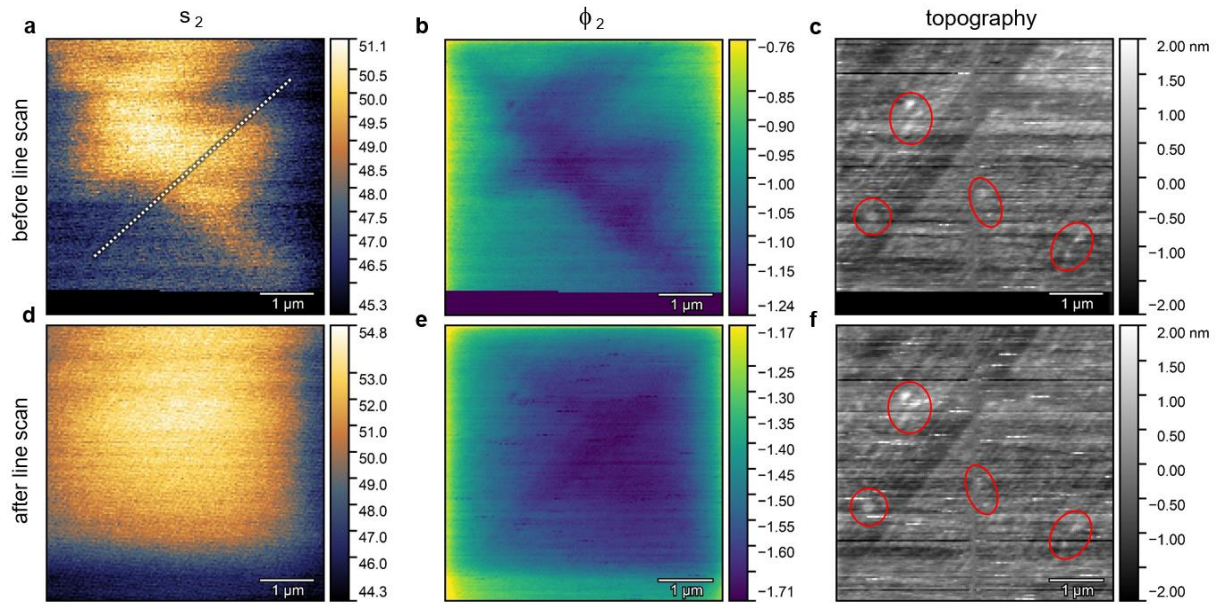

Figure S2: s-SNOM images of near-field amplitude $s_2$ (a, d), phase $\varphi_2$ (b,e) and topography (c,f), measured at $\omega = 700$ cm$^{-1}$. Subfigures a-c show the measurement before a hyperspectral line scan was recorded (partially shown in Figure 3b of the Main Text) along the dashed line indicated in Figure S2a. Subfigures d-f show a subsequent measurement, after the hyperspectral line scan was completed. Note that the first measurement (a-c) was aborted after 90% completion, indicated by a dark region at the bottom of each image. Red ovals in the topography highlight characteristic points, showing that position on the sample and scan range are identical.